\documentclass[twocolumn,aps,prl,floatfix,groupedaddress,longbibliography]{revtex4-1}

\usepackage{xcolor}
\usepackage[utf8]{inputenc} 
\usepackage{bm}
\usepackage{amsmath,amsthm,amssymb}
\usepackage{graphicx} 
\usepackage{pgfplots}
\usepackage{tikz,tkz-tab}{{}}
\usepackage[
  colorlinks=true,
  urlcolor=blue,
  linkcolor=blue,
  citecolor=blue
]{hyperref}
\usepackage[normalem]{ulem}
\usepackage{stackrel}
\usepackage{siunitx}

\def\urlprefix{}
\def\url#1{}

\long\def\soutjd#1{}

\newcommand{\bs}{\boldsymbol}

\begin{document}

\title{From many-body oscillations to thermalization in an isolated spinor gas}

\author{Bertrand Evrard, An Qu, Jean Dalibard and Fabrice Gerbier}

\affiliation{Laboratoire Kastler Brossel, Coll{\`e}ge de France, CNRS, ENS-PSL Research University, Sorbonne Universit{\'e}, 11 Place Marcelin Berthelot, 75005 Paris, France}

\date{\today}
\pacs{}

\begin{abstract}
The dynamics of a many-body system can take many forms, from a purely reversible evolution to fast thermalization. Here we show experimentally and numerically that an assembly of spin 1 atoms all in the same spatial mode allows one to explore this wide variety of behaviors. When the system can be described by a Bogoliubov analysis, the relevant energy spectrum is linear and leads to undamped oscillations of many-body observables. Outside this regime, the non-linearity of the spectrum leads to irreversibity, characterized by a  universal behavior. When the integrability of the Hamiltonian is broken, a chaotic dynamics emerges and leads to thermalization, in agreement with the Eigenstate Thermalization Hypothesis paradigm.
\end{abstract}

\maketitle


\paragraph{Introduction.} 

The temporal evolution of an isolated quantum system is governed by Hamiltonian dynamics and is in principle reversible. However, this apparently mundane statement masks a large range of possible scenarios, depending on the system size, its integrability (or lack thereof), the number of degrees of freedom, and also the observables under consideration. In practice, reversibility is observed for two-level systems or harmonic oscillators, for which the dynamics is time-periodic. When several Bohr frequencies contribute significantly to the evolution, one observes a rapid relaxation of initial oscillations, possibly followed by partial revivals. The system then reaches a quasi-stationary state that may be described by a Generalized Gibbs Ensemble (GGE), which takes into account the constants of motion on the considered time scale\,\cite{Jaynes:1957_PhysRev.106.620}. Finally, for sufficiently complex many-body systems, the Eigenstate Thermal Hypothesis (ETH) states that almost all initial conditions evolve towards a state leading to predictions indistinguishable from those of a thermal density matrix, at least for few-body observables\,\cite{dalessio2016,deutsch2018eigenstate,ueda2020quantum}.

To explore this variety of scenarios, multiple experimental platforms are usually required. Restricting for simplicity to atomic and photonic systems, the time-periodic behavior is typically observed using the resonant drive of an atomic transition with a coherent radio-frequency or light field\,\cite{haroche2006exploring}. The collapse and revival phenomenon was evidenced with an electromagnetic cavity containing a few photons\,\cite{Brune:1996} and with a few atoms trapped at the same site of an optical lattice\,\cite{Greiner:2002b}. Moving to larger systems for which the notion of GGE is relevant, relaxation dynamics was probed in several experiments with cold atoms\,\cite{gring2012,trotzky2012,langen2015} or with trapped ions\,\cite{Neyenhuis2017}. Thermalization was demonstrated to occur in small atom chains\,\cite{kaufman2016}, in dipolar atomic gases\,\cite{Tang:2018,lepoutre2019}, and in superconducting circuits\,\cite{Neill2016}. ETH was tested numerically for interacting particles on a lattice (see \textit{e.g.}\,\cite{kollath2007,rigol2008thermalization,dalessio2016} and references therein). Generally, ETH is expected to hold for chaotic systems with a large number of degrees of freedom. When facing this diversity of platforms, a natural question arises whether there exists a single system where all scenarios can be explored simply by tuning a few control parameters.   

In this Letter, we study experimentally and numerically the dynamics of a collection of $N$ spin-1 sodium atoms (with $N$ from $100$ to $5000$), all prepared in the same spatial mode in a tight laser trap. We show that the whole range of temporal behaviors mentioned above is accessible for the spin degrees of freedom. Since all atoms share the same spatial wave function, interactions between atoms are described by the Hamiltonian (up to an additive constant)\,\cite{ohmi1998,ho1998,yi2002,kawaguchi2012} 
\begin{align}
	\hat{H}_{\rm int}=\frac{U_s}{2N}\sum_{i,j=1}^N\hat{\bs s}_i\cdot\hat{\bs s}_j=\frac{U_s}{2N}\hat{\bs S}^2\,,\label{eq.Hamiltonian_int}
\end{align}
where $\hat{\bs s}_i$ is the spin of atom $i$, $\hat{\bs S}=\sum \hat{\bs s}_i$ is the total spin and $U_s/N$ is the spin-spin interaction strength. For sodium atoms, $U_s>0$ corresponds to antiferromagnetic interactions. In particular, the Hamiltonian $\hat{H}_{\rm int}$ describes the elastic two-atom spin-mixing process
\begin{equation}
(m=0)\ +\ (m=0)\ \leftrightarrows\ (m=+1) \ + (m=-1)\,,
\label{eq:collision_process}
\end{equation}
where $m=0,\pm 1$ is the quantum number associated with the component $\hat s_z$ of the spin of a single atom. This process, analog to optical parametric conversion\,\cite{walls}, has been used to generate entangled states\,\cite{duan2002,bookjans2011,bookjans2011,lucke2011,hamley2012,luo2017,zou2018,kunkel2018,fadel2018,lange2018,qu2020}. Here, together with a suitable one-body term, it generates a wide diversity of scenarios, from quasi-pure oscillations to thermalizing dynamics.


\paragraph{Many-body oscillations.}
In our setup, the atoms are immersed in a magnetic field ${\bs B}$ aligned along $z$, which shifts the energies of the $|m\rangle$ states. At first order in $B$, the Zeeman shift is proportional to $\hat{S}_z=\hat{N}_{+1}-\hat{N}_{-1}$, where $N_m$ is the number of atoms in state $|m\rangle$. It is a conserved quantity since $[\hat{S}_z,\hat{H}_{\rm int}]=0$, and thus does not contribute to the dynamics. For the relatively small field regime explored here, the relevant term is the quadratic Zeeman shift, which raises by $q\propto {\bs B}^2$ the energy of $|m=\pm 1\rangle$ with respect to $|m=0\rangle$. This leads to a Hamiltonian\,\cite{kawaguchi2012}
\begin{align}
	\hat{H}=\hat{H}_{\rm int}+ q\left(\hat{N}_{+1}+\hat N_{-1}\right)\,.\label{eq.Hamiltonian}
\end{align}

We start from the situation where each atom is in the Zeeman state $|m=0\rangle$. In this paragraph, we assume that $N_0$ remains large compared to $N_{\pm 1}$ at all times. In the spirit of the Bogoliubov approach for a scalar Bose gas, we treat the creation ($\hat a_0^\dagger$) and annihilation ($\hat a_0$) operators for the $|m=0\rangle$ state as c-numbers $\approx \sqrt N$ in the second-quantized expression of $\hat{H}_{\rm int}$. We are  left with a Hamiltonian quadratic with respect to the creation and annihilation operators in the weakly populated states $|m=\pm 1\rangle$:
\begin{eqnarray}
   \hat{H}&\approx&U_s \left( \hat a_{+1}^\dagger \hat a_{-1}^\dagger + \hat a_{-1} \hat a_{+1}\right) \nonumber \\
   &+& (q+U_s) \left( \hat a_{+1}^\dagger \hat a_{+1} + \hat a_{-1}^\dagger \hat a_{-1}\right).
\end{eqnarray}
It can be diagonalized using the Bogoliubov method\,\cite{landau1980statistical}, and one finds a linear spectrum of frequency\,\cite{kawaguchi2012,mias2008,SM}
\begin{equation}
	\hbar\omega_B=\sqrt{q(q+2U_s)}\,.\label{eq.BogoFreq}
\end{equation}

In the Bogoliubov regime, the dynamics is reversible and the mean number of pairs $(+1,-1)$ varies as\,\cite{mias2008,cui2008a}
\begin{align}
	\bar N_p(t)=\frac{U_s^2}{\hbar^2\omega_B^2}\sin^2(\omega_Bt)\,.\label{eq.NpBogo}
\end{align}
More precisely, the system is predicted to periodically evolve into a two-mode squeezed vacuum state, where the number of pairs follows the Bose-Einstein distribution\,\cite{mias2008,walls},
\begin{align}
	\mathcal{P}(N_p)\simeq\frac{1}{\bar N_p}\exp\left(-\frac{N_p}{\bar N_p}\right)\,,\label{eq.BogoDistrib}
\end{align}
with the standard deviation $\Delta N_p\simeq \bar N_p$. Note that the self-consistency of the approximation of the undepleted $m=0$ state requires $\bar N_p \ll N$, hence $q \gg U_s/N$.

\begin{figure}
	\centering
	\includegraphics[width=\columnwidth]{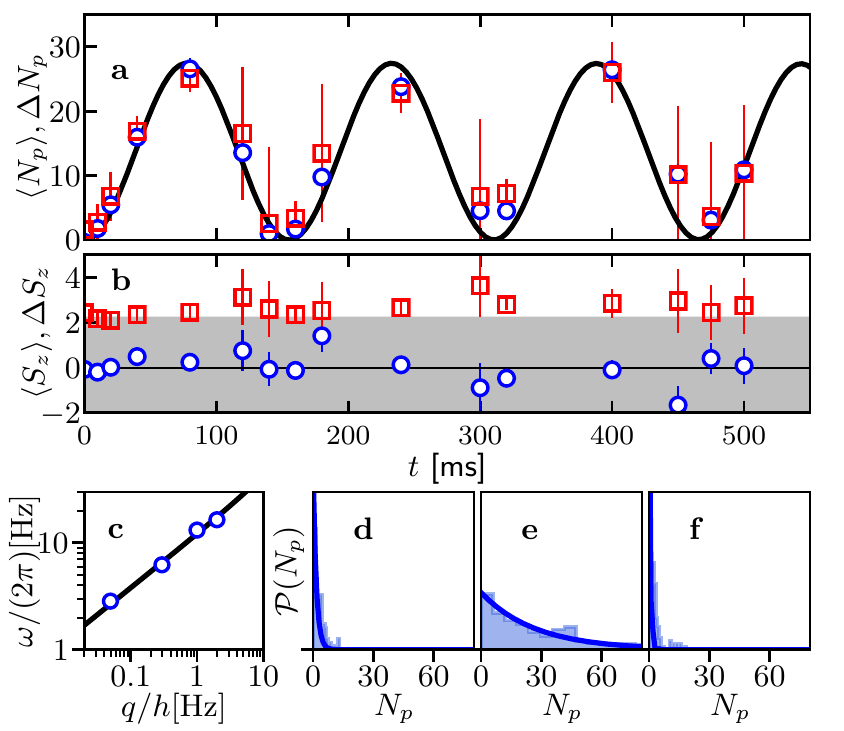}
	\caption{\textbf{Experimental observation of many-body oscillations.} ({\bf a}): Evolution of the mean number of pairs (circles)  and its standard deviation (squares) in a sudden quench from $q_i=277\,(3)\,$Hz to $q_f=0.31\,(1)\,$Hz. Here $N\approx5400\,(740)$. The solid line  is the result of a fit of the Bogoliubov prediction (\ref{eq.NpBogo}) to the experimental data, with $U_s=17.5\,(1.4)\,$Hz as a single  free parameter. ({\bf b}): Mean (circles) and standard deviation (squares) of the magnetization $S_z$. The shaded region indicates the detection noise level. ({\bf c}) Oscillation frequency obtained from a fit to the measured $\bar N_{\rm p}(t)$ (circles) for various $q$, compared to the prediction (\ref{eq.BogoFreq}) (line). 
	({\bf d}-{\bf f}) Measured distribution of the number of pairs for $t=10,\,80,\,160\,$ms, with the solid lines corresponding to the prediction (\ref{eq.BogoDistrib}).}\label{fig.1}
\end{figure}

To test this prediction, we prepared a Bose--Einstein condensate (BEC) with $N\sim 5000$ atoms in $|m=0\rangle$. We performed standard evaporative cooling in a field of $\sim 1\,$G in the presence of a magnetic force that removes all atoms in $|m=\pm1\rangle$. The magnetic field was then suddenly quenched to a lower value ($34\,$mG) to trigger the spin-mixing dynamics. Finally, we measured the populations $N_m$ in the three Zeeman states using fluorescence imaging, with a detection noise  $\Delta N_m\approx 1.6\,$ atoms\,\cite{qu2020}. 

We show in Fig.\,\ref{fig.1}ab the evolution of the mean value and standard deviation of $N_p=(N_{+1}+N_{-1})/2$ and $S_z=N_{+1}-N_{-1}$. We observe the predicted oscillations of $\bar N_p$ and verify that $\Delta N_p$ is almost equal to $\bar N_p$, as expected for a Bose distribution. The solid line in Fig.\,\ref{fig.1}a shows a fit of the expression (\ref{eq.NpBogo}) to the data, with $U_s$ as the only fit parameter. We also varied $q$ keeping $U_s$ constant and verified the prediction (\ref{eq.BogoFreq}) in Fig.\,\ref{fig.1}c. The magnetization $S_z$ remains compatible with zero at all times given our experimental resolution (fig.\,\ref{fig.1}b), which confirms that $m=\pm1$ atoms are produced in pairs. The non-classical character of the spin state can be inferred from the squeezing parameter $\zeta_s^2=\Delta \hat S_z^2/(2\bar N_p)$ \cite{qu2020,Vitagliano:2014_PhysRevA.89.032307,lucke2014}. At $t=80\,$ms we have $\bar N_p\approx 26.6$, $\Delta \hat S_z\approx2.45$ and $\zeta_s^2\approx0.11$ ($9.5$\,dB).

In Fig.\,\ref{fig.1}d-f, we investigate the evolution of the distribution of $N_p$ and show that it is well reproduced by a Bose distribution. It broadens for the first $80\,$ms, which could  be interpreted naively as a growth of entropy. However, this evolution is subsequently reversed almost perfectly and the systems returns  close to its initial state after $160\,$ms. 

The observation of beyond mean-field reversible evolution in a closed many-body system is an important result of this Letter. For comparison, a partial reversal of time evolution was achieved in a dynamically unstable BEC with modulated interactions\,\cite{hu2019}, using a sudden change of the relative phase between the various modes of the system. Combinations of closed, beyond mean-field evolution in an unstable regime and externally-driven rephasing sequences are also at the core of the SU(1,1) interferometers demonstrated in spinor BECs\,\cite{linnemann2016,linnemann2017,yurke1986}. In contrast, our experiment was performed with a stable system ($\omega_B$ real and positive) and no action was needed to reverse the dynamics. The isolated character of our system is essential for the subsequent discussion of relaxation and thermalization. 


\paragraph{Relaxation and Generalized Gibbs Ensemble.}

The oscillating behavior discussed above relies on the linearity of the  many-body spectrum $\sim n\hbar \omega_B$ ($n$ integer) in the Bogoliubov approximation. Outside this regime, the spectrum exhibits a significant non-linearity and the sum over several oscillation functions causes dephasing, as for the prethermalization phenomenon\,\cite{Berges:2004}. The expectation value of a physical observable relaxes to a steady-state value, possibly accompanied by revivals at some specific times. The spin-1 atomic assembly at zero magnetic field is well suited to observe such a behavior since the spectrum of $\hat{H}_{\rm int}$ is $E_S=S(S+1)U_s/2N$, hence quadratic with the quantum number $S$ associated with the total spin\,\cite{law1998,ho2000,koashi2000}. In practice, the  magnetic field should be such that $q \ll U_s/N$ to ensure that the Zeeman energy is negligible for the states that we consider hereafter. 

We first investigate theoretically the relaxation associated with this quadratic spectrum. We consider again the initial state $|\psi_i\rangle=|m=0\rangle^{\otimes N}$ and study its evolution for a zero magnetic field. For $N\gg 1$, the decomposition of $|\psi_i\rangle$ on the basis states $|S,M\rangle$, where $M$ is the quantum number associated with $\hat S_z$,   reads\,\cite{diener2006a}
\begin{equation}
|\psi_i\rangle =\sum_S c_S\,|S,0\rangle, \quad c_S\approx \sqrt{\frac{2S}{N}}{\rm e}^{-S^2/4N}.
\label{eq:initial_state}
\end{equation}
Here, the sum runs on even (resp.\,odd) values of $S$ for $N$ even (resp.\,odd) and the most populated spin states are $S\sim \sqrt{N}$. Using the matrix elements of $\hat N_0$ between spin states for $S\ll N$,
\begin{equation}
\langle S,0|\hat N_0|S',0\rangle\approx \frac{N}{2}\delta_{S,S'}+\frac{N}{4}\left(\delta_{S,S'-2}+\delta_{S,S'+2}\right), 
\label{eq:matrix_element}
\end{equation} 
and treating $S$ as a continuous variable, we find that the evolution of the population $n_0=\langle N_0\rangle/N$ obeys
\begin{equation}
n_0(t) = 1 -  \tau D(\tau),\qquad \tau=\sqrt{\frac{2}{N}}\,\frac{U_s t}{\hbar},
\label{eq:relax_univ}
\end{equation} 
where  $D(\tau)=\int_0^{+\infty} \sin(2x\tau)\, {\rm e}^{-x^2}\,{\rm d }x$ is the Dawson function. At long times, $n_0(t)$ tends to $1/2$.

\begin{figure}[t]
\begin{center}
\includegraphics[width=8cm]{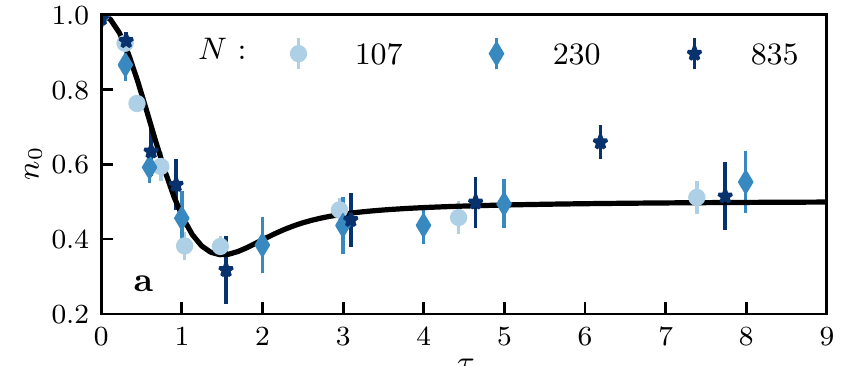}
\includegraphics[width=8cm]{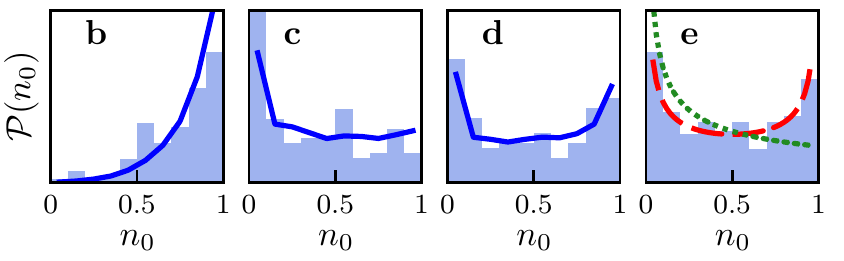}
\end{center}
\caption{\textbf{Experimental observation of relaxation near zero magnetic field.} (a) Evolution of the population $n_0(t)$ following a fast quench of $q$ to a negligible value, for various atom numbers $N$. Disks: $N=107$, $U_s=17.2\,$Hz; squares: $N=230$, $U_s=24.2\,$Hz; Losanges: $N=835$, $U_s=64.7\,$Hz. The initial state is $|m=0\rangle^{\otimes N}$. For $\tau=9$, the ``real" time spanned is $t=609$\,ms, $635$\,ms, and $452$\,ms for the three atom numbers. The solid line is the universal prediction (\ref{eq:relax_univ}). (b-e) Distribution of the population $n_0$ at $t=30, 100, 200,500$\,ms for $N=107$ atoms. In (b-d) the solid lines are the results of a numerical simulation. In (e) the green dotted line is the prediction from the micro-canonical ensemble\,\cite{Tasaki:2013_PhysRevLett.110.230402} and the red dashed line is the prediction from the GGE with the constraint $S_z=0$\,\cite{SM}.}
\label{fig:relax_univ}
\end{figure}

We now turn to the experimental investigation of this relaxation dynamics, with atom numbers in the range $100$--$1000$. The spin interaction was calibrated using the oscillations of $N_{\pm 1}$ at relatively large $q$ (see Fig.\,\ref{fig.1}) and ranges from $U_s/h=17\,$Hz for $N=110$ up to $50\,$Hz for $N=840$. We performed a sudden quench to $q=11\,$mHz ($B=6.2$\,mG) so that the inequality $Nq < U_s$ is well satisfied for all atom numbers. We show in Fig.\,\ref{fig:relax_univ} the evolution of  $n_0$. The agreement with the prediction (\ref{eq:relax_univ}) is excellent. In particular, the collapse of data acquired with notably different atom numbers shows that the relaxation dynamics is entirely characterized by the ``universal" function $\tau D(\tau)$. We checked for all data of Fig.\,\ref{fig:relax_univ}a that the magnetization $S_z$ remains  compatible with zero, as in Fig.\,\ref{fig.1}c.
  
Fig.\,\ref{fig:relax_univ}a shows no sign of revival, neither for the experimental data, nor for the theoretical prediction (\ref{eq:relax_univ}). The lack of revival in the theory is an artifact of the replacement of the discrete sum over $S$ in Eq.\,(\ref{eq:initial_state}) by an integral. Keeping $S$ as a discrete quantum number, the time-dependent phase factors ${\rm e}^{-{\rm i}E_St/\hbar}$ appearing in $|\psi(t)\rangle$ rephase at times multiple of $hN/U_s$\,\cite{diener2006a}. In practice, this time is much larger than 1\,s even for our lowest atom number, and parasitic effects such as atom losses prevented us to observe these revivals.

In spite of the SO(3) symmetry of $\hat H_{\rm int}$, the three populations $n_m$ are not all equal to $1/3$ after relaxation is complete. This implies that the final state of the spin assembly cannot be described by a thermal density matrix such as $\hat \rho \propto \exp(-\hat H_{\rm int}/k_{\rm B}T)$ associated with an effective temperature $T$. This non-thermal character is readily explained by the fact that $S_z$ is a conserved quantity. For a given $S_z$, the system has essentially a single degree of freedom characterized by the value of the quantum number $S$ and is thus integrable\,\cite{Lamacraft:2011}.    

In order to apply the statistical physics formalism to such a case, one has to consider a GGE that takes into account the conservation of magnetization\,\cite{Jaynes:1957_PhysRev.106.620,dalessio2016,rigol2007relaxation,kinoshita2006,langen2015}. For our choice of initial state, with mean energy $\bar E=\langle \psi_i|\hat H_{\rm int}|\psi_i\rangle\approx U_s$ and vanishing average magnetization, the density matrix associated with this GGE is 
\begin{align}
\hat \rho_{\rm GGE}\propto\sum_{E_S\in {\cal W}} |S,0\rangle\langle S,0| .\label{eq.GGE}
\end{align}
Here the sum runs over the spin states $|S,0\rangle$ whose energy $E_S$ sits in a narrow window ${\cal W}$ centered on $\bar E$. From the matrix elements given in Eq.\,(\ref{eq:matrix_element}), one deduces that the GGE average population $\bar n_0=\frac{1}{N}{\mbox Tr}(\hat \rho_{\rm GGE} \hat N_0)=1/2$, which coincides with the asymptotic result predicted in Eq.\,(\ref{eq:relax_univ}) and measured experimentally. 

To compare more precisely our results with the predictions of the GGE, we consider the distribution ${\cal P}(N_0)$  plotted in Fig.\,\ref{fig:relax_univ}b-e at four different times for $N=107$. We observe that this distribution reaches a steady-state value in excellent agreement with the one calculated with $\hat \rho_{\rm GGE}$, plotted as a dashed line in Fig.\,\ref{fig:relax_univ}e. To the contrary, the prediction for ``true" thermal equilibrium, which is obtained by extending the sum (\ref{eq.GGE}) to all spin states $|S,M\rangle$ in the energy window ${\cal W}$, differs significantly from the experimental result (dotted line in Fig.\,\ref{fig:relax_univ}e).    

In practice, the relaxation of macroscopic observables in spinor BECs may also originate from fluctuating initial states or from couplings between  spin and spatial modes  \cite{kronjaeger2005a,kronjaeger2006a,Tian2020,Yang2019,evrard2019,liu2009}. 
Here, these processes are negligible compared to the self-relaxation due to the non-linearity of the energy spectrum and characterized by the universal law (\ref{eq:relax_univ}). 


\paragraph{Chaotic dynamics and thermalization.}

For a generic many-body system, the Eigenstate Thermalization Hypothesis (a term coined in \cite{srednicki1994}) states that essentially any energy eigenstate $|\psi_E\rangle$ is ``typical", in the sense that the statistical properties of a few-body observable $\hat {\cal O}$ evaluated with $|\psi_E\rangle$ are close to their expectation value for thermal equilibrium, calculated for example using the micro-canonical density matrix  $\hat \rho_E$ at the energy $E$ (for a review, see \cite{dalessio2016,deutsch2018eigenstate,ueda2020quantum}). A consequence of ETH is thermalization: If we consider an initial wave packet $|\psi(t=0)\rangle$ formed by a combination of many energy eigenstates all around the energy $E$, the time average of $\langle \psi(t)|\hat {\cal O}|\psi(t)\rangle$ will be close to the thermal equilibrium average ${\rm Tr}(\hat \rho_E\hat{\cal O})$.

The justification of ETH is closely related to the theory of random matrices\,\cite{deutsch1991,srednicki1994,Brody_1981:RevModPhys.53.385}. More precisely, the validity of ETH is established for systems with a large number of degrees of freedom whose level statistics corresponds to a chaotic behavior such as the spectrum of random matrices from the grand orthogonal or grand unitary ensembles. On the contrary, when the level statistics corresponds to a regular motion, ETH does not hold. 

To address the connection between thermalization and chaos for a spin 1 ensemble, we consider the following Hamiltonian:
\begin{equation}
\hat H'=\hat H + \Omega \hat S_x\,.
\label{eq:H_prime}
\end{equation}
The second term in $\hat H'$ breaks the integrability of  $\hat{H}$ by ensuring that $\hat S_z$ is not conserved anymore, so that the two degrees of freedom associated with the quantum numbers $S$ and $S_z$ are now coupled. The one-body term $\Omega \hat S_x$ describes (in the rotating frame) the effect of a resonant  coupling induced by a rotating radio-frequency field that drives the transitions $|m\rangle \leftrightarrow |m\pm 1\rangle$. Note that the implementation of $\hat H'$ requires an excellent control of the ambiant magnetic field in order to keep the fluctuations of the first-order Zeeman effect small compared to $U_s$ \footnote{For $U_s/h\approx 20\,$Hz, the field noise must be $\Delta B\ll30\,\si{\micro G}$, a much more stringent condition than the constraint $\delta q\ll U_s/N$ ($\Delta B\ll30\,$mG), for the experiment shown in Fig.\,\ref{fig:relax_univ}. Although it is beyond the performance of our machine, such a stability can be achieved with a specifically designed spinor gas setup\,\cite{farolfi2019design}.}.

The matrix of $\hat H'$ in the particle number basis $|N_{-1},N_0,N_{+1}\rangle$ (with $N_{\pm 1}=(N-N_0\pm M)/2$) is real and symmetric. We diagonalized it numerically for $N=100$ and studied its level statistics as a function of the control parameters $U_s$, $q$ and $\Omega$. While the spectrum corresponds to a regular motion when one of the three parameters is either large or small compared to the two others, a chaotic behavior emerges when they are all comparable \cite{Rautenberg2020}.

\begin{figure}[t]
\begin{center}
\includegraphics{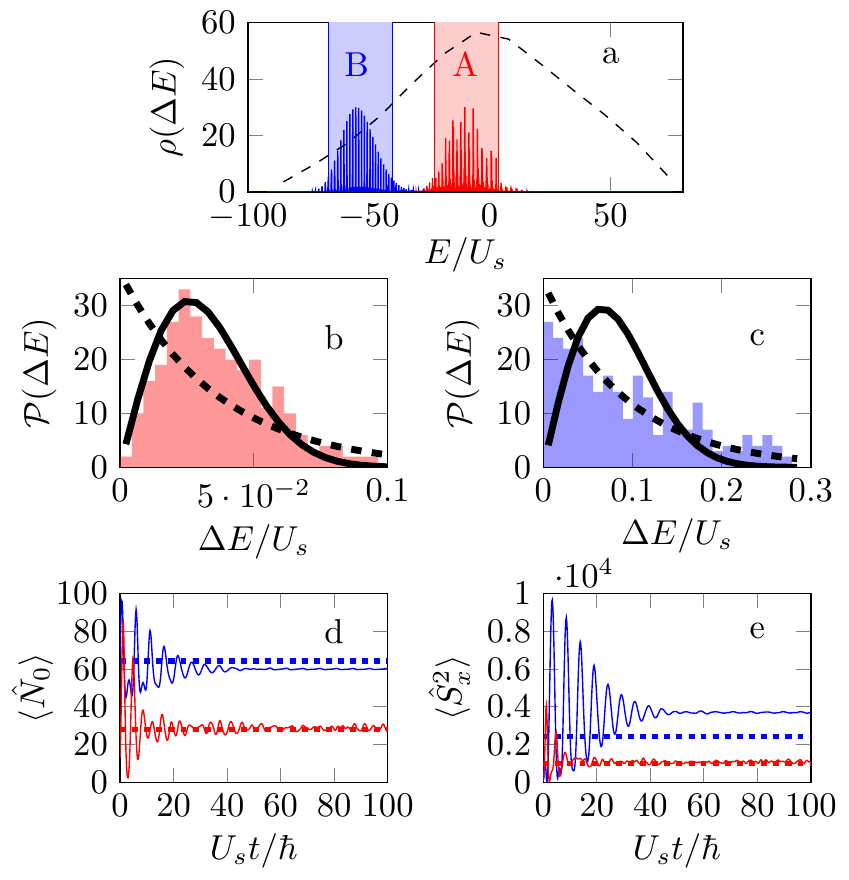}
\end{center}
\caption{\textbf{Numerical exploration of chaotic behavior and thermalization.} (a) Density of states (dashed line) for the Hamiltonian $\hat H'$ for  $q/U_s=0.8$, $\Omega/U_s=0.6$ and $N=100$. The histograms show the populations (in arbitrary units) of the states chosen in (d-e) to investigate the dynamics in the colored zones $A$ and $B$. (b-c)  Statistics of the splitting $\Delta E$ between adjacent energy levels for zones $A$ and $B$, containing respectively 724 and 286 levels. The continuous and dashed lines show the  Wigner--Dyson (chaotic motion) and Poisson (regular motion) distributions, respectively \cite{bohigas1984characterization}. (d-e) Evolution of  $\langle \hat N_0\rangle$ and $\langle\hat S_x^2\rangle$ for the wave packet localized in zone $A$ or $B$ shown in (a). The dashed lines are the predictions from the micro-canonical ensemble.}
\label{fig:chaos}
\end{figure}

We show in Fig.\,\ref{fig:chaos}a the density of states calculated for $q/U_s=0.8$ and $\Omega/U_s=0.6$. In Fig.\,\ref{fig:chaos}bc we give the distributions of the splittings between adjacent energy levels, for eigenstates inside the two shaded zones of Fig.\,\ref{fig:chaos}a. In zone A (in red, centered on $E/U_s=-10$), adjacent levels show a clear antibunching, close to the one expected for random real symmetric matrices,  an indicator of chaotic dynamics\,\cite{bohigas1984characterization}. On the contrary for zone $B$ (in blue, centered on $E/U_s=-53$), the level statistics is close to a Poisson law, characteristic of a regular motion. 
This diagnostic is confirmed by a resolution of the mean-field equations, which shows chaotic (respectively regular) trajectories, for initial conditions corresponding to an energy in zone A (resp. B)\cite{SM}. Another chaotic feature of $\hat H'$, the growth of out-of-time-ordered correlators at short times, was also evidenced in \cite{Rautenberg2020}.

To analyze the relationship between the emergence of a chaotic behavior and ETH for our spin system, we calculated the evolution of various physical quantities for a wave packet inside zone $A$ or $B$. The initial state is $\hat{\cal R}_x(\theta)|m=0\rangle^{\otimes N}$, where $\hat{\cal R}_x(\theta)$ is the rotation around the $x$ axis, with the angle $\theta$ adjusted such that the average energy of the state sits in the middle of the desired zone.
We show in Figs.\,\ref{fig:chaos}d-e the evolution of the expectation values of the one-body observable $\hat N_0$ and the two-body observable $\hat S_x^2$, together with the thermal averaged value of these quantities, using the micro-canonical density matrices for zones $A$ and $B$. The results fully confirm the prediction of the ETH: for the wave packet prepared inside the chaotic region ($A$), full thermalization does occur whereas for the wave packet in region $B$,  the asymptotic value of $\langle \hat S_x^2\rangle$ differs significantly from the thermodynamic average. In addition, relaxation is notably faster in the chaotic sector than in the regular one, a hierarchy that also occurs in the model  developed in \cite{garcia2018} for a quasi one-dimensional Bose gas.   


To summarize, a collection of spin 1 atoms in the same spatial mode allows one to study the large variety of time-evolutions that are accessible to an isolated many-body system. Depending on the applied magnetic field, we could observe either a beyond mean-field oscillating dynamics, or an irreversible relaxation characterized by a universal function. We also showed that this system is well suited to test the ETH, as it provides the minimal number of degrees of freedom for the emergence of a chaotic behavior in a closed system.

\vskip 2mm
\begin{acknowledgments}	
\noindent  We thank Isabelle Bouchoule, Markus Oberthaler and Maxim Olshanii for insightful discussions and comments. This work was supported by the European Research Council (Synergy Grant UQUAM). LKB is a member of the network ``Science and Engineering  for Quantum Technologies in the Ile-de-France Region".
\end{acknowledgments}

\bibliographystyle{apsrev}

\end{document}